\documentclass[aps,prb,showpacs,groupedaddress,twocolumn]{revtex4}
\usepackage{graphicx}
\usepackage{verbatim}
\usepackage{mathrsfs}
\usepackage{subfigure}
 \usepackage{gensymb}
\usepackage{color}
\usepackage{natbib}
\def\be{\begin{equation}}
\def\ee{\end{equation}}
\def\bea{\begin{eqnarray}}
\def\eea{\end{eqnarray}}
\def\bw{\begin{widetext}}
\def\ew{\end{widetext}}
\def\nn{\nonumber}

\usepackage{amsmath,amsfonts,amssymb}

\def\3{2.8in}    
\def\2{2.5in}
\def\4{3.0in}

\begin{document}
\title{Simplified model on the timing of easing the lockdown}
\author{Sung-Po Chao}
\affiliation{Department of Physics, National Kaohsiung Normal University, Kaohsiung 82444, Taiwan, R.O.C.}
\date{\today}
\begin{abstract}
Lockdown procedures have been proven successful in mitigating the spread of the viruses in this COVID-19 pandemic, but they also have devastating impact on the economy. We use a modified Susceptible-Infectious-Recovered-Deceased  model with time dependent infection rate to simulate how the infection is spread under lockdown. The economic cost due to the loss of workforce and incurred medical expenses is evaluated with a simple model. We find the best strategy, meaning the smallest economic cost for the entire course of the pandemic, is to keep the strict lockdown as long as possible.
\end{abstract}
\maketitle

\section{Introduction}
The COVID-19 pandemic is an ongoing global pandemic of corona virus disease (COVID) first identified in Wuhan, China, in December 2019\cite{wikic}. As of 20 July 2020, more than 14.6 million confirmed cases of COVID-19 have been reported globally\cite{news}. The virus is mainly spread between healthy and infected people during close contact, most often through small droplets produced by coughing, sneezing, and talking\cite{wikic}. Thus maintaining proper physical distance from other people, keeping good indoor ventilation, and wearing a face mask\cite{fmask} in public places should be effective\cite{mask2} in reducing the spread of the virus. People may also get infected by touching a contaminated surface and then touching their face. Regularly disinfecting surfaces in public places and keeping the habit of hand washing for everyone should also curb COVID-19 infection.\\ 

Lockdown refers to the adoption of strict rules imposed by the local government to decrease individual mobility and increase physical distancing, sometimes accompanied by mandatory use of masks or other personal protective equipment in public places. Based on the studies of mobile phone tracking done in China\cite{MUG} and Italy\cite{Marco}, the effectiveness of lockdown in controlling the spread of the virus depends on how strict the lockdown procedures are implemented. With a tighter lockdown, sufficient low mobility does bring down transmission below the level needed to sustain the epidemic\cite{Marco}. As vaccine against COVID-19 is currently not available\cite{NYTracker}, lockdown remains to be one of the proven effective approach\cite{cnnr} for curbing the spread of infection, especially if contact tracing\cite{SK,Chen} were not likely to be done efficiently. However these lockdown measures also come with many serious economic impacts, especially for the low income households\cite{Martin1}, along with some social and psychological tensions within the populations\cite{Salah,Rahman}. How to strike the balance between controlling the spread of the virus and maintain social economic stability becomes the key issue for navigating out the crisis brought by this pandemic. Some countries, such as Sweden and South Korea, have not fully locked down\cite{busi}, and its success\cite{svsk} depends on efficient contact tracing and massive tests to identify and quarantine those who are infected. Frequent random testing\cite{Markus} may also help to avoid complete lockdowns. In the event of major outbreak, the large amount of infected people would make contact tracing ineffective against the spread of infection\cite{lsc}, and some forms of locked down seems to be the most effective approach to curb the spread of infection.\\

The goal of this paper is to find the optimal timing, in the sense of lowest economic cost for the entire course of the pandemic, for easing the strict lockdown. Less stringent controls are required following the end of the strict lockdown to temporarily end the pandemic. We model the pandemic using the simplest compartment model\cite{compart} called Susceptible-Infectious-Recovered-Deceased (SIRD) model with time dependent infection rate to simulate the effect of lockdown. The economic cost during the pandemic is vastly simplified to include only two parts: the economic cost due to the loss of available workforce and lockdown regulations, and the medical expenses for the infected. Based on this model study, in a three-stages scenario staring with natural spread, followed by strict lockdown, and finally less stringent lockdown, the best strategy is to keep the strict lockdown as long as possible. Three countries, U.S.A., Italy, and China, are selected to see if indeed the longest strict lockdown, which is indeed good for temporarily stopping the pandemic, is also good for the economy. The changes in their quarterly GDP\cite{ecodata} seems to suggest that is correct, but no clear indication suggests this from their unemployment rate data\cite{UEM}. More detailed comparisons in different economic indexes should be made.\\

There are several existing literatures discussing economic trade-offs and optimal policy analysis within the SIR framework\cite{Martin,Fernando,Jones,Maryam,Daron,Lukasz,Toda,Erhan}, or its extension such as SEIR\cite{Ellina,Saket} (E stands for "exposed"), or other more sophisticated compartment models\cite{Andrea,David}. Other than these so called equations based approach, agent based model of COVID-19 epidemic\cite{Silva,Kano,Dignum,Todo}, which takes more computational efforts, is also applied on these issues. A nice relevant literature review can be found in Ref.~\onlinecite{Andrea}. Most of the studies support strict and long lockdown as advocated by this paper. The main difference between this work and others is that the end of pandemic defined here (for evaluation of economic cost during the pandemic) is not the real ending of the pandemic. That is, the herd immunity is not discussed in this model study. The reason is that, without sufficient study on the duration of immunity and reliable vaccines for this virus, it would inevitably cost many more lives, as briefly discussed in section \ref{simsc}, to achieve the herd immunity. New lifestyles of epidemic prevention\cite{tw} should be followed during the post pandemic time.\\

This paper is organized in the following way. In the section \ref{ms}, we introduce a slightly modified SIRD model and a simple model for economic cost due to the pandemic and lockdown. In the section \ref{simsc}, economic costs for different duration of lockdown are evaluated with simplified time dependent infection rate. Results are summarized in the Table \ref{tab1}. More realist time dependent infection rate for three countries (U.S.A., Italy, and China; data collected up to July 9th, 2020) and some of the associated economic cost are shown in the section \ref{three}. In the section \ref{conclusion} we summarize our results, and comment on the limitations of the studies done in this paper. \\
\section{Model discussion}\label{ms}
To model the pandemic for some number of population $N$, we use a slightly modified SIRD model\cite{note1}, incorporating the loss of lives when the infection rate goes above some "medical threshold". The equations describing the ratio of susceptible $S(t)$, currently infected $I(t)$, the recovered ones $R(t)$, and the deceased ones $D(t)$ are given by:
\bea\nn
 \frac{dS(t)}{dt}&=&-\beta(t) S(t) I(t) ,\\\nn
\frac{dI(t)}{dt}&=&\beta(t) S(t) I(t)-\gamma(t) I(t)\\\nn
&-&\eta(t)I(t)\Theta(I(t)-I_m) ,\\\nn
 \frac{dR(t)}{dt}&=&\gamma(t) I(t) ,\\\label{SIR model}
 \frac{dD(t)}{dt}&=&\eta(t)I(t)\Theta(I(t)-I_m).
\eea
 Here $\beta(t)$ is a time dependent infection rate, which decreases from some maximal value set by the nature of the virus to positive number close to zero if very strict lockdowns were imposed. $\gamma(t)$ is the recovery rate given by the average time it takes for the infected one to recover from this disease. $\eta(t)$ is a time dependent death rate, and $\Theta(x)$ is the Heaviside step function being $1$ when $x\ge 0$ and $0$ otherwise. Parameter $I_m$ can be evaluated by the fatality rate in the region we discuss, and is related to the medical resources available in that region during this pandemic.\\ 
 
 We use a naive model to describe the economic impact brought by the lockdown. Gross domestic product per capita at nominal values or gross domestic product per capita at purchasing power parity of the country/region, labeled indiscriminately as $GDP$ hereafter, is treated as the average economic output contributed by a healthy person (which includes susceptible and recovered ones). The influence of the lockdown procedures on the economic output is modeled by a productivity decrease
function $L_\alpha(t)$ with $0 \le L_\alpha(t) \le 1$. That is, $L_\alpha(t)\times GDP \times N(S(t)+R(t))$ is the economic output contributed by the healthy people within the total population $N$ at some time $t$. $L_\alpha(t)$ can be modeled by some empirical economic data, and it should vary significantly for different type of work or region we are interested in. For example, the tourism, restaurant, and public transport are probably strongly influenced by the lockdown, while the business related to online transactions such as stock trading are much less influenced. Here we model $L_\alpha(t)=(\beta(t)/\beta(0))^{\alpha}$ by treating this economic output, on average, relying on the human to human contact as in the SIRD model with $0\le \alpha\le1$. $\alpha=0$ is used for the business pattern where no actual human contact is required (such as online lectures), and $\alpha=1$ is used for the business pattern where human contact is crucial (such as traditional factories, tourisms, cultural performances, etc.).\\  

At the same time, the economic cost for taking care of the infected is given by $ME\times N I(t)$ where $ME$ stands for the average medical expenses per person. The economic cost at time $t$ due to the lockdown and the additional cost on the medical treatment is then given by: 
\bea\label{eco}
GDP \times N \Big(1-L(t)\big(S(t)+R(t)\big)\Big)+ME\times N I(t)
\eea
From Eq.(\ref{eco}) we note that the crucial factor is $ME/GDP$ as the value of $GDP$ and $N$ are assumed to be fixed (time independent) in this model. If $ME/GDP\gg 1$ then it would be wiser to keep the infection rate $I(t)$ as low as possible, at the expense of slower lifting of the lockdown. If $ME/GDP\simeq 1$ then an early lifting of the lockdown (which does not mean going completely back to the usual normal life!) might be better for the economic benefit of general public. \\

Denote "$T$" as the amount of time for stopping the first wave of the pandemic (defined here as $I(T)\simeq I(0)$, the initial infection ratio, without clear signature of resurgence of the pandemic after some time), and $t=0$ is the start of this pandemic. In between we may have two special points in this time frame: the time to start the lockdown, and the time for easing or lifting of the lockdown, denoted as $s_lT$ and $e_lT$ respectively. It follows that $0< s_l<e_l\le 1$.\\ 

\section{Simplified scenarios}\label{simsc}
For the simplified scenarios, we treat the influence of imposing or lifting the lockdown on the $\beta(t)$ by an abrupt change in its value. Other parameters are simplified to be time independent, with $\gamma(t)=\gamma=0.1$, $\eta(t)=\eta=0.03$, and $I_m=5\times 10^{-5}$. During the hard (strict) lockdown the $\beta(t)$ is fixed at $\ln(1.02)$, and for the post lockdown period the $\beta(t)$ is fixed at $\ln(1.09)$. Before the lockdown is imposed, the $\beta(t)$ is fixed at $\ln(1.4)$, corresponding to the natural spreading rate of the virus. The "end" of this (first wave of) pandemic is defined as the ratio of the active cases lowered to its initial value, which is set to be $I(0)=10^{-6}$. \\

\begin{figure}
  \includegraphics[width=\linewidth]{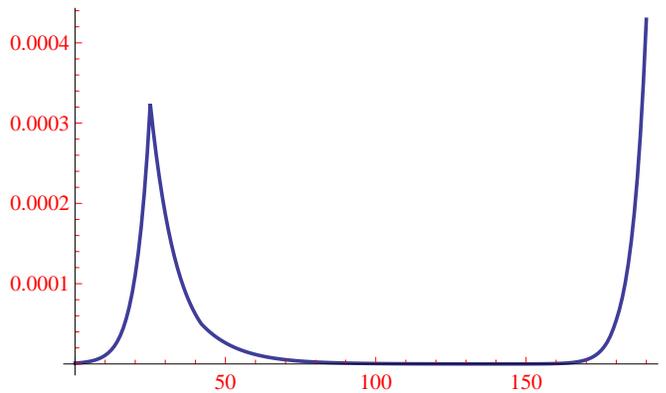}
  \caption{Daily active case ratio $I(t)$ as a function of time $t$ (in the unit of day). $\beta(t)=\ln(1.4)$ for $0\le t\le 25$ and $145<t$, and $\beta(t)=\ln(1.02)$ for $25<t\le 145$. $I(t)$ drops below $I(0)=10^{-6}$ after $t=90$ but raises again significantly after $t=165$. This figure is shown for time $t$ up to $t=190$.}
  \label{fig0}
\end{figure}

This, however, does not really mean the life can go back to completely normal as setting the $\beta(t)$ back to $\ln(1.4)$ at arbitrary later time will cause a second wave of major outbreak. One such example is shown in the Fig.\ref{fig0} with the second outbreak occurs after around $20$ days of going back to completely normal. This demonstrates the importance of new lifestyles of epidemic prevention without herd immunity during the post pandemic time. These new lifestyles include, but not limited to, the wearing of face masks at the crowded public places, washing hands regularly, border quarantine, and quick response to trace and isolate possible sources of community infection. All these efforts are made to control the $\beta(t)$ to be below a critical value $\gamma(t)/S(t)$, so that there is no second wave of the pandemic. In the model parameters mentioned here, this critical $\beta(t)\simeq \ln(1.105)$ since the majority of the population are not infected. That is, the susceptible ratio $S(t)$ is still close to $1$ (the largest active case ratio, or the peak of $I(t)$ in the Fig.\ref{fig0} is $3.3\times 10^{-4}$ during the first pandemic).\\  

The only exception to the above statement is when herd immunity is achieved when a large portion of a community becomes immune to the COVID-19. Without available vaccination, this herd immunity could possibly be achieved when nothing is done to prevent the spread of the infection. For the parameters given above with $\beta(t)$ fixed at $\ln(1.4)$ all the time, we find this pandemic officially ends at $t=223$ (no possible second wave). On this day $20.83\%$ of the population is dead due to this virus, with about $9.64\%$ of the population still susceptible and about $69.53\%$ of the population recovered with immunity. Apparently this high ratio of mortality is not what we want, and some forms of lockdown or quarantine measures are needed before effective vaccine is available to lower the number of deaths due to this virus.\\   

To evaluate the economic cost due to the influence of the pandemic, we integrate Eq.(\ref{eco}) with respect to time (with three stage $\beta(t)$ mentioned in the last paragraph) and set $GDP\times N$ to be $1$ for simplicity. Different $\alpha$ are used for $L(t)$ in the Eq.(\ref{eco}) to simulate the different impacts on different business. In the Table \ref{tab1}, we listed $\alpha=1$, $0.1$, and $0.01$ cases with three different strict lockdown time $(e_l-s_l)T=30$, $45$, and $60$ days. We choose $ME/GDP=13$ in this calculation, but we shall bear in mind that this number could vary significantly even over different regions in the same country. As losses of human life are very personal and difficult to estimate economically, we choose $I_m=5\times 10^{-5}$ on purpose in this simplified scenarios to let the mortality rate in  
all three cases to be the same. The mortality rate is defined as the total number of deaths divided by the population $N$ after the pandemic is over, or
\bea\nn 
\text{mortality rate}\equiv 1-\big(S(T)+I(T)+R(T)\big)=D(T)
\eea  

\begin{figure}
  \includegraphics[width=\linewidth]{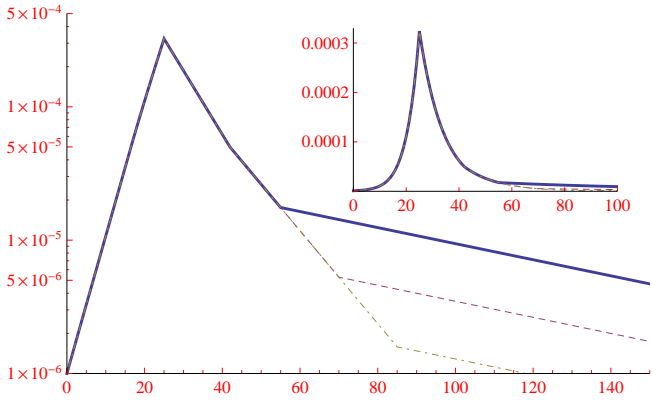}
  \caption{Active cases ratio $I(t)$ versus time $t$. Inset is plotted in the linear scale while the main figure is plotted with semi-log scale. Blue-thick line is for $30$ days of strict lockdown, purple-dashed line is for $45$ days of strict lockdown, and brown-dot dashed line is for $60$ days of strict lockdown.}
  \label{fig2}
\end{figure}

The choice of $I_m$ can be understood from Fig.\ref{fig2}, which shows the daily active rate $I(t)$ as a function of time $t$. The inset shows the original (linear) scale, and the main figure is plotted with the semi-log scale which helps to illustrate the exponential growth/suppression more clearly. $I(t)$ is suppressed more strongly for $I(t)>5\times 10^{-5}$ (the slope in the semi-log, or $d\ln(I(t))/dt$, is more negative). For $t>45$ (or after $20$ days of strict luckdown) $I(t)$ has already dropped below $I_m$, and therefore in all three cases (with all $t>55$) we discuss here the mortality rate is the same.\\

In the Table \ref{tab1}, we list the ending day of the pandemic $T$, the economic cost from day $1$ to $T$ (labeled as $E_{cost}^{Total}$), and the economic cost from day $1$ to $t=90$ (labeled as $E_{cost}^{90 days}$ with three different $\alpha$ values mentioned above. The mortality rate, shown at the bottom of the table, in all three cases are kept at $0.01\%$. The fatality rate, defined as number of deaths divided by those who have been infected, at the end of the pandemic, is given by
\bea\nn
\text{fatality rate}\equiv\frac{\Big(1-\big(S(T)+I(T)+R(T)\big)\Big)}{1-S(T)}=\frac{D(T)}{1-S(T)}
\eea
 The fatality rate is listed in the 5th row of the Table \ref{tab1}. From the row atop the bottom of Table \ref{tab1}, we see that longer lockdown makes the fatality rate slightly higher (around $18\%$ for 60 days lockdown compared with $15\%$ for 30 days). This is because the total number of people infected is larger for shorter lockdown with fixed number of deaths in our model.\\

From the second row of Table \ref{tab1}, we see that the end of this first wave of pandemic depends crucially on how long this strict lockdown lasts. Two months of strict lockdown (with $T=119$ days) leads to the "end of the tunnel" after $34$ days of easing the lockdown, while
one month of strict lockdown (with $T=262$ days) needs additional $143$ days (or $4$ more months) to see this. In fact, for $65$ days of strict lockdown the active cases ratio $I(t)$ drops below $10^{-6}$ as mentioned in the caption of the Fig.\ref{fig0}. This shows that longer strict lockdown effectively shortens the total time for the pandemic and decreases the total number of infected people.\\  

\begin{table}[h!]
\centering
 \begin{tabular}{||c|| c| c| c||} 
 \hline
 $(e_l-s_l)T$ & 30 & 45 & 60 \\
 \hline
 $T$ & 262 & 191 & 119 \\
 \hline
 $\alpha=1$ & 181.57 & 131.70 & 81.84 \\

 $E_{cost}^{Total}$ $\alpha=10^{-1}$ & 33.73 & 26.47 & 19.21 \\ 

 $\alpha=10^{-2}$ & 3.73 & 2.96 & 2.21 \\
 \hline
  $\alpha=1$ & 54.35 & 57.31 & 60.27 \\

 $E_{cost}^{90 days}$ $\alpha=10^{-1}$ & 11.93 & 13.72 & 15.51 \\ 

 $\alpha=10^{-2}$ & 1.39 & 1.60 & 1.81 \\
 \hline
  Fatality rate & $15.21\%$ & $17.12\%$ & $17.79\%$ \\
 \hline
 Mortality rate & $0.01\%$ & $0.01\%$ & $0.01\%$\\
 \hline
\end{tabular}
\caption{Table for simplified scenarios. $GDP\times N=1$ and $ME/GDP=13$ are used in this calculation. Unit of time is "day". Three different $\alpha$ values are used to illustrate different impact on different businesses from the pandemic. The economic cost is evaluated numerically by integrating Eq.(\ref{eco}) with respect to time from $t=0$ to $t=T$ for $E_{cost}^{Total}$ or $t=0$ to $t=90$ for $E_{cost}^{90days}$. Changing $ME/GDP$ to $1$ (not shown in the table) only slightly reduces the economic cost and does not change the overall pattern (as $I(t)$ considered here is very small).}
\label{tab1}
\end{table}
The economic cost till the end of the pandemic and after 90 days since the beginning of the pandemic are shown respectively in the third and forth row of the Table \ref{tab1}. For a given $(e_l-s_l)T$, both $E_{cost}^{Total}$ and $E_{cost}^{90days}$ show that $\alpha=10^{-2}$, i.e. the business least relying on the actual human contact in the three cases we discuss here, has the lowest economic cost (impact from the pandemic) as expected. For a given $\alpha$, $E_{cost}^{90days}$ is lowest for $(e_l-s_l)T=30$ and largest for $(e_l-s_l)T=60$. The  $E_{cost}^{Total}$ shows completely opposite trend, being lowest for $(e_l-s_l)T=60$ and largest for $(e_l-s_l)T=30$ with a large differences in economic cost compared with $E_{cost}^{90days}$ for a given $\alpha$. These results show that the \emph{short strict lockdown is indeed good for the economy in the short term, but bad for the economy in the long run} in our simple model. This statement is true for all businesses, as represented by three different $\alpha$ here, having negative impact (or positive economic cost here) from the pandemic.\\

Based on the simplified scenarios, the best strategy for reducing the economic impact seems to be prolonging the strict lockdown until the end of the pandemic. However, aforementioned analysis does not include the problem of bankruptcy for the businesses severely impacted by the lockdown. For example, the largest economic cost $E_{cost}^{max}$, for some company modeled by $\alpha=1$, is set to be $E_{cost}^{max}=55$. For economic cost beyond this value, the company would go bankruptcy and causing unemployment problems. So the employers and/or employees of the company would surely favor lifting the strict lockdown as soon as possible. This yearning for shorter lockdown is explained in the following paragraph.\\

 In the simplified scenarios, the company can survive in the short term (i.e. $E_{cost}^{max}>E_{cost}^{90days}$) if the strict lockdown is short (say, $(e_l-s_l)T=30$). It goes bankruptcy for longer lockdown (such as $(e_l-s_l)T=45,60$) within 90 days. Thus the owners or some employees of the company would favor lifting the strict lockdown as early as possible, without knowing that shorter lockdown actually brings even more cost in the long run. That is, $E_{cost}^{max}<E_{cost}^{T'}$ for some $90<T'<T$ even for $(e_l-s_l)T=30$ case. The best strategy for the company then is to keep the strict lockdown as long as it can, and how long the company may survive under the lockdown crucially depends on $E_{cost}^{max}$. To have longer survival time, the company needs to increase $E_{cost}^{max}$ with the help from banks (lowering interests rate) and timely government subsides, and to modify the business patterns to lower its $\alpha$ value (changing the in person meetings to online ones, for example). Some passenger airplanes having been modified to serve as cargo planes in some airlines is one good example of lowering
the $\alpha$ value in the airline business.\\

In short, "the more haste, the less speed" or "haste makes waste" accurately depicts the shortcomings of the hasty lifting of the lockdown. During this pandemic we shall
"overcome impetuosity and exercise patience, and go steady so we can go far". Let us use the modified SIRD model described in the Eq.(\ref{SIR model}) to study more realistic cases, and compare the lockdown procedures and timings between these cases in the following section. The methods used to obtain
the parameters $\beta(t)$, $\gamma(t)$, $\eta(t)$, and $I_m$ in this modified SIRD model are introduced in the beginning of the next section.

\section{More realistic cases}\label{three}
To compare with the actual cases, we evaluate the $\tilde{\beta}(t)$ by fitting the actual data nicely organized by Wade Fagen-Ulmschneider\cite{91Codata}. $\tilde{\beta}(t)$ is a time dependent infection rate in the modified susceptible-infected (SI) model with equations given by:
\bea\nn
 \frac{dS(t)}{dt}&=&-\tilde{\beta}(t) S(t) \tilde{I}(t) ,\\\label{SI model}
\frac{d\tilde{I}(t)}{dt}&=&\tilde{\beta}(t) S(t) \tilde{I}(t),
\eea
\begin{figure}[h!]
  \includegraphics[width=\linewidth]{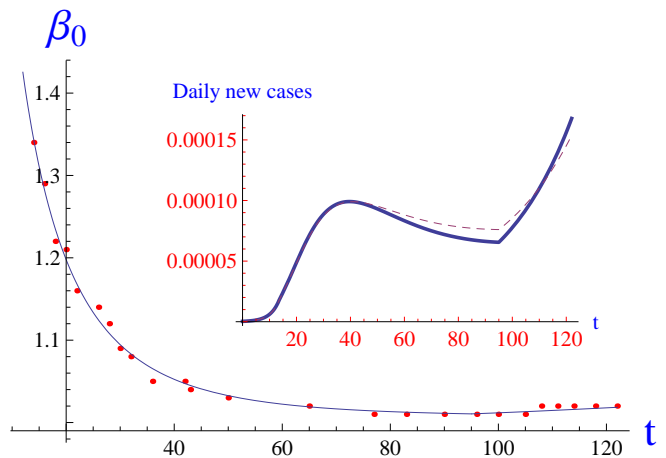}
  \caption{Geometric growth number $\beta_0(t)$ v.s. time $t$ for the U.S.A. case. Red circular dots are the data taken from Ref.~\onlinecite{91Codata} and thin blue line is part of the fitting function $\beta_f(t)$. Before $t=14$, $\beta_0(t)$ is roughly fixed at $1.35$, which is fitted by a constant value (not shown here) in that region. Inset shows daily new cases v.s. time.}
  \label{USfig1}
\end{figure}
and the susceptible ratio $S(t)$ and infected ratio $\tilde{I}(t)$ are constrained by $S(t)+\tilde{I}(t)=1$. Three countries, U.S.A., Italy, and China, are chosen to be presented here as they are the three countries with the first major outbreak in America, Europe, and Asia, respectively. The duration of the data taken is from days since 1 case per million people till July 9th, 2020, which is 122 days for U.S.A., 136 days for Italy, and 163 days for China.

\subsection*{U.S.A. case}
 The actual data points used to obtain $\tilde{\beta}(t)$ is shown as red circular dots in the Fig.\ref{USfig1}, in which the data from the U.S.A. is used as an example of how we use the actual data. Those points represent the geometric growth number $\beta_0(t)$ of a given day obtained from the data of total confirmed cases shown in the Ref.~\onlinecite{91Codata}. We fit $\beta_0(t)$ with some mathematical function $\beta_f(t)$, and part of  the fitting function $\beta_f(t)$ (after $t=14$, as before $t=14$ the $\beta_0(t)\simeq 1.35$ is roughly a constant) is shown as a thin blue line in the Fig. \ref{USfig1}. For the U.S.A. case, the $\beta_f(t)$ is:
\bea\nn
&&\beta_f(t)=1.35\Theta(14-t)+\Theta(t-14)(1.007356+6.159892\\\label{bf}
&&\times e^{-0.7773997\sqrt{t}}+0.01(t-95)\Theta(t-95)/27)
\eea

 Time dependent infection rate $\tilde{\beta}(t)$ in the SI model is related to the fitted geometric growth function $\beta_f(t)$ via $\tilde{\beta}(t)=\ln(\beta_f(t))$. Using this $\tilde{\beta}(t)$ we solve SI model numerically to obtain the daily new cases $\frac{d\tilde{I}(t)}{dt}$. The result is shown as the thick blue line in the inset of Fig.\ref{USfig1}, which is roughly consistent with the new cases shown in Ref.~\onlinecite{91Codata}.\\

\begin{figure}
  \includegraphics[width=\linewidth]{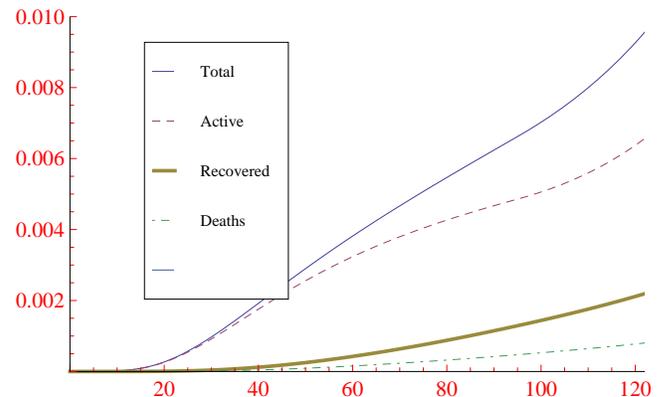}
  \caption{Linear scale for the U.S.A. cases v.s. time (up to July 9th,2020). The total confirmed case, the active cases, the recovered, and the deaths ratio are plotted as thin blue line, dashed purple line, thick brown line, and dot dashed green line}
  \label{USfig2}
\end{figure}
\begin{figure}
  \includegraphics[width=\linewidth]{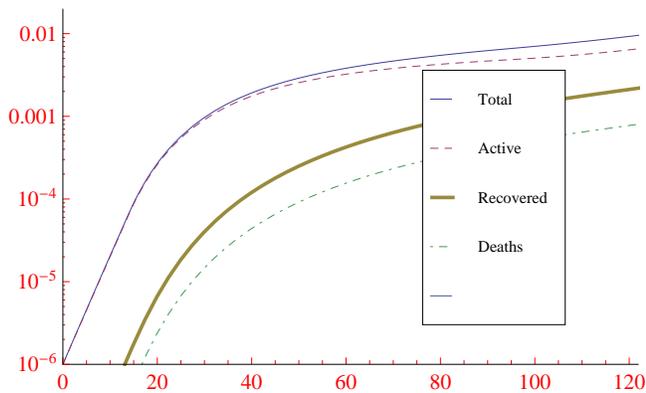}
  \caption{Semi-logarithmic scale for the U.S.A. cases v.s. time (up to July 9th,2020). For Fig.\ref{USfig3} and Fig.\ref{ITfig2} to Fig.\ref{Chfig3}, we use the same color codes as those used in the Fig.\ref{USfig2}.}
  \label{USfig3}
\end{figure}

 To obtain the parameters for the modified SIRD model as described in the Eq.(\ref{SIR model}), we add a small positive constant on $\tilde{\beta}(t)$ to get $\beta(t)$, and choose other parameters to be time independent, i.e. $\gamma(t)=\gamma$, $\eta(t)=\eta$. In the U.S.A. case presented here, $\beta(t)-\tilde{\beta}(t)=0.0053$ is chosen, so that the daily new cases obtained in this modified SIRD model, plotted as dashed purple line in the inset of Fig. \ref{USfig1}, is roughly consistent with that obtained from the SI model (blue thin line in the same figure). The parameters $\gamma$, $\eta$, and $I_m$ are chosen by requiring the total confirmed case, the active (or currently infected) cases, the recovered, and the deceased ratio to be roughly consistent\cite{note3} with the actual data at the last day ($t=122$ in this U.S.A. case) of the data set.\\ 
 
 Following aforementioned methods, we find that $\beta(t)-\tilde{\beta}(t)=0.0053$, $\gamma=0.006$, $\eta=0.0022$, and $I_m=1\times 10^{-5}$ giving rise to the total confirmed case, the active cases, the recovered, and the deaths ratio for the U.S.A. up to July 9th. These computed results are plotted as thin blue line, dashed purple line, thick brown line, and dot dashed green line in the Fig.\ref{USfig2} and Fig.\ref{USfig3}. These two figures show the same results with different scale (linear for Fig.\ref{USfig2} and logarithmic for Fig. \ref{USfig3}) on the vertical axes. The logarithmic scale plot Fig.\ref{USfig3} shows more clearly the onset of recovered and death ratio, which is around 13 and 17 days after the onset of the active cases respectively. The actual data\cite{91Codata} happens to be the reverse of the two, i.e. the onset of recovered ratio is around 17 days and that of death ratio is around 13 days. This discrepancy reflects that the actual death rate $\eta$ and recovery rate $\gamma$ are time dependent\cite{note3}, possibly reflecting the shortage of sufficient medical resources in the beginning of the pandemic. As we mainly focus on 
the timing of lifting the strict lockdown in this article, we stick with the choice of parameters similar to the simplified scenarios mentioned in the last section. That is, only $\beta(t)$ is time dependent and all other parameters in the modified SIRD model are time independent.\\

\begin{figure}
  \includegraphics[width=\linewidth]{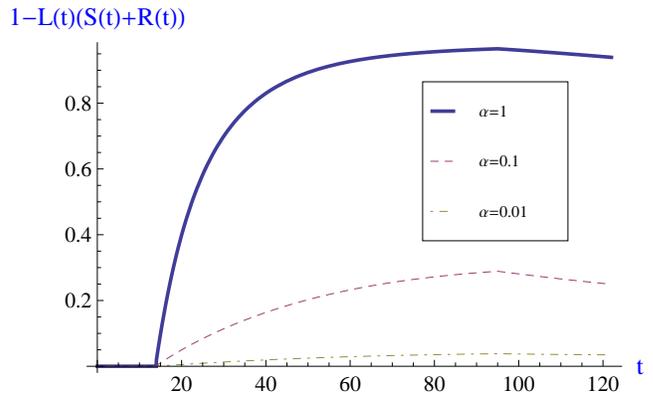}
  \caption{Economic cost $1-L(t)(S(t)+R(t))$ v.s. time $t$ for $\alpha=1$ (blue thick line), $\alpha=10^{-1}$ (purple dashed line), and $\alpha=10^{-2}$ (brown dot dashed line). Medical cost is not included here as the comparison is done with the monthly unemployment rate.}
  \label{USfig4}
\end{figure}

Based on the report of the New York Times\cite{NYT}, there is no national-wide lockdown imposed by the federal government\cite{BI} in the U.S.A., perhaps due to some legal issues\cite{MSN}. Many states did have imposed some shelter-in-place rules or state-wide stay at home order starting in late March ($t\simeq 20$ in Fig.\ref{USfig1}-\ref{USfig3}), affecting $94\%$ of the country's population at the peak of the lockdown based on the report of Business Insider\cite{BI}. Starting around the end of April, many states had lifted the lockdown or restrictions with multiple stages. For the states taking more cautious steps for reopening, such as New York and New Jersey (both with major reopening around late May/early June, or around $t\simeq 80$ in Fig.\ref{USfig1}-\ref{USfig3}), have successfully reduced the spread of the coronavirus and flatten the epidemic curve, while for some states like Arizona, Florida, and Louisiana, have seen rapid increase in the daily new confirmed cases\cite{91Codata}. This effectively changes the curve of the national-wide daily new cases, as shown in the inset of Fig.\ref{USfig1}, and prolongs the duration of this COVID-19 pandemic.\\

One aspect of the economic impact due to this pandemic can be judged by the monthly unemployment rate\cite{UEM}. The unemployment rate increases from $3.5\%$ on February 20 to $4.4\%$ on March 20 ($t\simeq 10$), and jumps to $14.7\%$ on April 20 ($t\simeq 40$). After the lifting of lockdown in late April, the unemployment rate drops down continuously to $13.3\%$ in May and $11.1\%$ in June. The sudden increase of monthly unemployment rate is roughly captured by the economic cost $1-L(t)(S(t)+R(t))$ with $\alpha=1$ case, as shown in the Fig.\ref{USfig4}. This $\alpha=1$ curve reaches maximum at $t\simeq 95$ (as is expected from the $\beta_f(t)$ in Eq.(\ref{bf})), so the simple economic cost function proposed here does not give the correct timing for the drop in unemployment rate.\\

 For Italy and China, the monthly unemployment rate is not completely available on the Internet. So we focus on how the timing/procedures of imposing/easing the lockdown influence their time dependent infection rate $\beta(t)$ and other relevant parameters in the modified SIRD model in the following. We comment briefly on the monthly unemployment rate from the influence of the pandemic for these three countries in the last part of this section.       

\subsection*{Italy case}
\begin{figure}[h!]
  \includegraphics[width=\linewidth]{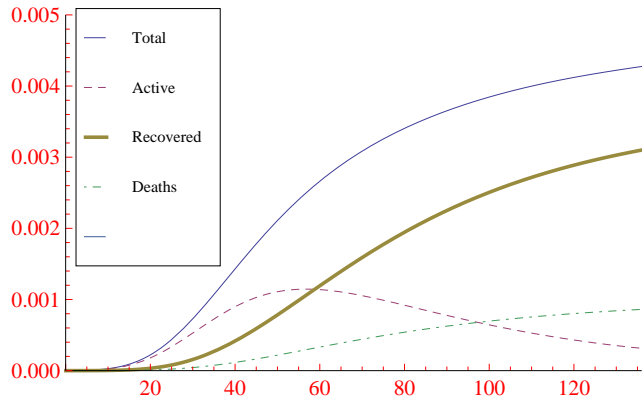}
  \caption{Linear scale for Italy cases v.s. time (up to July 9th,2020).}
  \label{ITfig2}
\end{figure}

\begin{figure}[h!]
  \includegraphics[width=\linewidth]{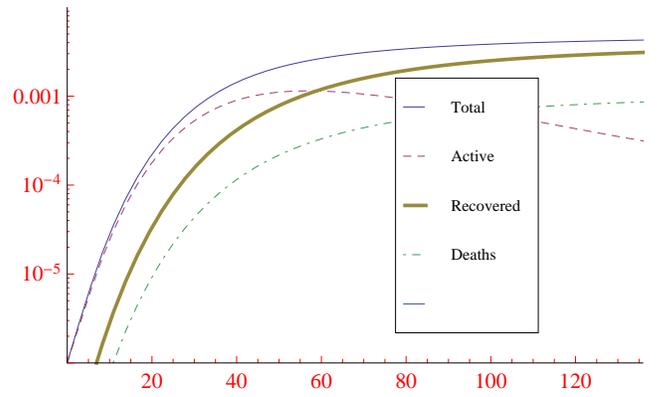}
  \caption{Semi-logarithmic scale for Italy cases v.s. time (up to July 9th,2020).}
  \label{ITfig3}
\end{figure}

Following the same methods as in the U.S.A. case, we find the fitted geometric growth function $\beta_f(t)$ for Italy is given by:
\bea\nn
\beta_f(t)&=&1.42\Theta(6-t)+\Theta(t-6)(0.996986\\\nn
&+&0.582248 e^{-0.05821783t}).
\eea
We choose $\beta(t)-\tilde{\beta}(t)=0.028$, $\gamma=0.036$, $\eta=0.01$, and $I_m=1.5\times 10^{-6}$ to match the total confirmed case, the active cases, the recovered, and the deaths ratio for Italy up to July 9th. These computed results are plotted in the Fig.\ref{ITfig2} and Fig.\ref{ITfig3}.  The logarithmic scale plot Fig.\ref{ITfig3} shows more clearly the onset of recovered and death ratio, which is around 7 and 10 days after the onset of the active cases respectively. This is very close to the actual data set, with recovered and death ratio around 8 and 10 days after the onset. 

The lockdown in Italy started from March 9th, and ends on May 18th\cite{ITwiki}. The starting and ending day corresponds to $t\simeq 20$ and $t\simeq 86$ in Fig.\ref{ITfig2} and Fig.\ref{ITfig3}. It started in Northern Italy, but later expanded to the whole of Italy a day later. The lockdown measures implemented by Italy was considered the most radical ones\cite{bbcn} except those implemented by China\cite{rfi}. The active cases has dropped down consistently from its peak value (at around $t=53$ in Fig.\ref{ITfig2}).\\

 Italy's current emergency rules remain in place until at least the end of July. The current measures include the obligation to wear face masks on public transport and in shops, restaurants, public offices, hospitals, and workplaces where it is not possible for people to keep at least one meter apart at all times\cite{itn}. With the current trend, the pandemic is expected to stop ($I(t)\simeq 10^{-6}$), without the help of vaccination, for around one year and three months from now.

\subsection*{China case}
\begin{figure}[h!]
  \includegraphics[width=\linewidth]{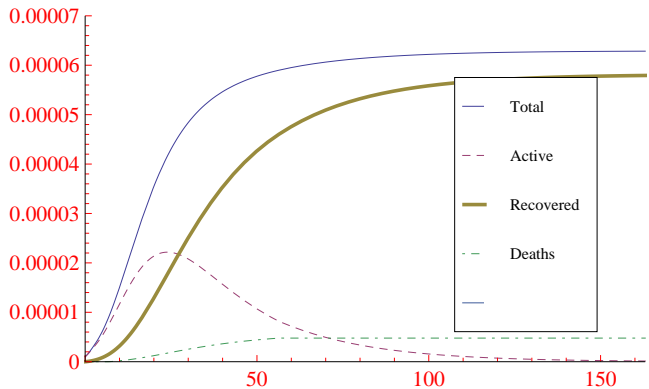}
  \caption{Linear scale for China cases v.s. time (up to July 9th,2020).}
  \label{Chfig2}
\end{figure}

\begin{figure}[h!]
  \includegraphics[width=\linewidth]{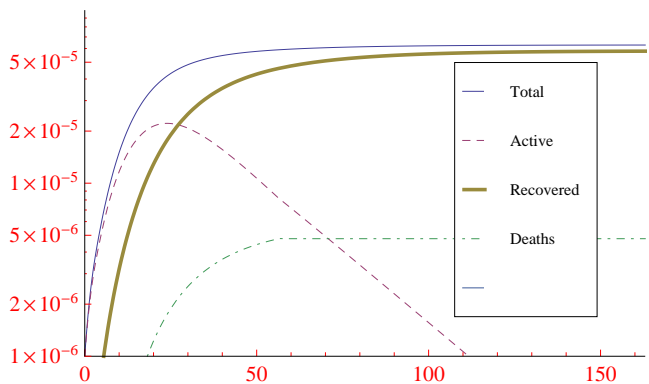}
  \caption{Semi-logarithmic scale for China cases v.s. time (up to July 9th,2020).}
  \label{Chfig3}
\end{figure}

Following the same methods as in the U.S.A. case, the fitted geometric growth function $\beta_f(t)$ for China is given by:
\bea\nn
&&\beta_f(t)=(0.6 (1 - t/4)^2 + 1.3)\Theta(4-t)+\Theta(t-4)\\\nn
&&(0.996561+0.449182 e^{-0.09589504t}+30.173546e^{-t^2}).
\eea
We choose $\beta(t)-\tilde{\beta}(t)=0.021$, $\gamma=0.056$, $\eta=0.006$, and $I_m=8\times 10^{-6}$ to match the total confirmed case, the active cases, the recovered, and the deaths ratio for China up to July 9th. These computed results are plotted in the Fig.\ref{Chfig2} and Fig.\ref{Chfig3}.  The logarithmic scale plot Fig.\ref{Chfig3} shows more clearly the onset of recovered and death ratio, which is around 6 and 19 days after the onset of the active cases respectively. This is roughly consistent with the actual data\cite{91Codata} with the onset of recovered and death ratio around 11 and 20 days after the onset of the active cases.\\


The lockdown in China started from January 23, with partial lifting on 19 March\cite{chwiki1}, and ends on April 8th for Wuhan\cite{chwiki2}. The starting and ending day corresponds to $t\simeq 1$ and $t\simeq 75$ in Fig.\ref{Chfig2} and Fig.\ref{Chfig3}. It started in Wuhan city in Hubei, China, but later expanded to many parts of China later. From Fig.\ref{Chfig3} we see that the pandemic stops at around $t=110$. The actual data\cite{91Codata} actually shows the pandemic stops at around $86$ days from the onset of the pandemic.\\

The reason why the lockdown in China was imposed on $t\simeq 1$ rather than $t\simeq 20$ as in Italy and U.S.A. might have to do with the Wuhan as the
ground zero (in the sense of the first city for large scale cluster infection) of coronavirus pandemic\cite{cnn}. The actual infected number might be much larger than the confirmed (reported) ones as the number of available testing initially was not sufficient. Also the majority of the cases were reported in Wuhan, and by the time of imposing the lockdown on $t=1$ the infected number is also around 100 people per million if we treat all reported cases in Wuhan and thus the denominator is the population of Wuhan (around 11 million) instead of China. This amount of population ratio is similar to the ones in the U.S.A. and Italy (c.f. Fig.\ref{USfig3} and Fig.\ref{ITfig3}).\\  

\subsection*{Comparison of three cases}
Aforementioned three cases happen to roughly illustrating three cases in the Table.\ref{tab1}. Among the three countries, China has the longest lockdown time and strictest lockdown regulations\cite{gua}. It turns out to be the most effective one for ending the pandemic, but the strict regulations may endanger the human rights\cite{alj} and pose negative psychological impact\cite{Dubey}.\\ 
 
The next question would be that if this longer, stricter lockdown does do good for the economy, as is suggested in the simplified scenarios. The answer to that is not clear, as we need more transparent data for making a fair comparison. The second quarterly (Q2) GDP\cite{ecodata} in China rebounded to $3.2\%$ growth from a record $6.8\%$ contraction in the previous (Q1) quarterly GDP (announced on April 17 for Q1 and July 16 for Q2). Q1 GDP in the U.S.A.\cite{ecodata} shrank $5\%$, following the $2.1\%$ growth in Q4 last year (announced on June 25 for Q1 and March 26 for last year's Q4). Q1 GDP in Italy\cite{ecodata} shrank $5.3\%$, following the $0.3\%$ contraction in Q4 last year (announced on May 29 for Q1 and March 4 for last year's Q4). The Trading Economics forecast\cite{ecodata} for Q2 of U.S.A. and Italy are $-27\%$ and $-13.2\%$, respectively. If the forecast were correct, then this trend in the quarterly GDP suggests longer, stricter lockdown might indeed be beneficial for the overall economy.

\section{Conclusion}\label{conclusion}
In this paper we treat the effect of lockdown as changing the values of time dependent infection rate $\beta(t)$ in the SIRD model described in the Eq.(\ref{SIR model}). Other parameters $\gamma(t)$ and $\eta(t)$ are kept to be time independent to simplify the calculation. Combining the obtained results with a naive economic cost function described in Eq.(\ref{eco}), we compute the economic cost for the entire course of lockdown and economic cost after 90 days (counting from the beginning of the pandemic) in simplified scenarios, where $\beta(t)$ changes discontinuously during different stages of lockdown. The results, summarized in Table.\ref{tab1}, suggest that long strict lockdown
is the most beneficial approach for economy in the long run, even with the same number of deaths due to the pandemic\cite{note4}. More realistic computations for $\beta(t)$ are done in the section.\ref{three} for U.S.A., Italy, and China, and their respective economic indicators during this pandemic are discussed. Quarterly GDP seems to suggest the conclusion reached in the simplified scenarios, but the monthly unemployment rate does not support that.\\

The epidemic model we adopted is the simplest compartment model\cite{compart}, so the proper comparison with the model predictions should be done at
smaller scale (such as county or city) with few external traffic flows (so that the population roughly maintains a constant) instead of countries mentioned in the section.\ref{three}. However, other than some states in the U.S.A., there are not many regional data/information available on the internet for this COVID-19 pandemic. Given the simplicity of the epidemic model mentioned in Eq.(\ref{SIR model}) and proposed phenomenological economic cost function in Eq.(\ref{eco}), it is easy to adapt the relevant parameters to determine, from data and the model prediction\cite{fauci}, when would be the proper time to ease the lockdowns\cite{note5}, with proper inter-city or inter-regional quarantine measures, at some local community. Another merit of Eq.(\ref{eco}) is that company facing financial crisis may predict the possible losses based on the current trend of its financial data, as the calculation done in the Fig.\ref{USfig4}, and ask for timely government/bank support.\\  
  
Last but not least, most of the compartment model studies, including this one, assumes that the recovered people are immune to the virus. However the concentration of associated antibodies seems to decrease with time after recovered from this COVID-19\cite{Seow}. This does not necessarily mean that 
long term immunity is not possible for COVID-19 as T cells may fight coronavirus in absence of antibodies\cite{tcell}. More time dependent studies on both the relevant antibodies and T cells shall shed light on the issue of how effective the vaccination would be. Another important issue not discussed in fixed population compartment model studies is about the international travel post the COVID-19. A combination of flow models\cite{Zlojutro,Zlojutro1} or network models\cite{Sullivan} with the compartment model can be applied to determine the optimal ways for easing the regulations on the international travel during the post pandemic time. 
   
\section*{Acknowledgment}
This work is done with the financial support from MOST in Taiwan (Grant No.106-2112-M017-002-MY3). I also thank useful discussion with Prof. Yu-Lin Chao from school of medicine, Tzu chi university near the final stage of this work.


\begin{thebibliography}{}
\bibitem{note3} We can also obtain all time dependent parameters empirically by setting $I_m=0$ (or some positive constant). $\beta(t)$ is determined from 1st line of Eq.(\ref{SIR model}) with input of daily new infection $-dS(t)/dt$, susceptible ratio $S(t)$, and active cases ratio $I(t)$. $\eta(t)$ is computed from 2nd line of Eq.(\ref{SIR model}) with input of daily deaths ratio and active cases ratio $I(t)$. $\gamma(t)$ is evaluated from 3rd line of Eq.(\ref{SIR model}) with input of daily recovered ratio $dR(t)/dt$ and active cases ratio $I(t)$.
\bibitem{note4} More realistically, shortening the strict lockdown prolongs the duration of the pandemic. This also leads to increase of number of deaths (as can be checked by requiring $I_m=0$), leading to even larger loss 
in the simple economic cost function described in the Eq.(\ref{eco}).
\bibitem{note5} Instead of choosing a particular holiday with no particularly good reasons, we may work it out with available data and some mathematics. This is the reason I started to work on this project near the beginning of April. At the same time I also convinced a group of undergraduate students to switch from solid state physics study club (Steve Simon's book) to the compartmental models in epidemiology. Their report will be published elsewhere.
\end{thebibliography}
\end{document}